\documentclass[final]{aipproc}
\layoutstyle{8x11double}

\newcommand{\etal}{et\,al.\ }
\newcommand{\logg}{\mbox{$\log g$}}
\newcommand{\Teff}{\mbox{$T_\mathrm{eff}$}}

\begin{document}

\title{Light and heavy metal abundances in hot central stars of planetary nebulae}

\classification{97.10.Ex}
\keywords      {stars: atmospheres --
                stars: evolution --
                stars: AGB and post-AGB --
                stars: white dwarfs --
                planetary nebulae: general}

\author{Klaus Werner}{
  address={Institut f\"ur Astronomie und Astrophysik, Universit\"at
  T\"ubingen, 72076 T\"ubingen, Germany}
}

\author{Agnes I.\,D. Hoffmann}{
  address={Institut f\"ur Astronomie und Astrophysik, Universit\"at
  T\"ubingen, 72076 T\"ubingen, Germany}
}

\author{Dorothee Jahn}{
  address={Institut f\"ur Astronomie und Astrophysik, Universit\"at
  T\"ubingen, 72076 T\"ubingen, Germany}
}

\author{Thomas Rauch}{
  address={Institut f\"ur Astronomie und Astrophysik, Universit\"at
  T\"ubingen, 72076 T\"ubingen, Germany}
}

\author{Elke~Reiff}{
  address={Institut f\"ur Astronomie und Astrophysik, Universit\"at
  T\"ubingen, 72076 T\"ubingen, Germany}
}

\author{Iris Traulsen}{
  address={Institut f\"ur Astronomie und Astrophysik, Universit\"at
  T\"ubingen, 72076 T\"ubingen, Germany}
}

\author{Jeffrey W. Kruk}{
  address={Department of Physics and Astronomy, Johns Hopkins University,
Baltimore, MD 21218, USA}
}

\author{Stefan Dreizler}{
  address={Institut f\"ur Astrophysik, Universit\"at
  G\"ottingen, Friedrich-Hund-Platz 1, 37077 G\"ottingen, Germany}
}

\begin{abstract}
We present new results from our spectral analyses of very hot central stars
achieved since the last IAU Symposium on planetary nebulae held in Canberra
2001. The analyses are mainly based on UV and far-UV spectroscopy performed with
the \emph{Hubble Space Telescope} and the \emph{Far Ultraviolet Spectroscopic
Explorer} but also on ground-based observations performed at the \emph{Very
Large Telescope} and other observatories. We report on temperature, gravity, and
abundance determinations for the CNO elements of hydrogen-rich central stars. In
many hydrogen-deficient central stars (spectral type PG1159) we discovered
particular neon and fluorine lines, which are observed for the very first time
in any astrophysical object. Their analysis strongly confirms the idea that
these stars exhibit intershell matter as a consequence of a late helium-shell flash.
\end{abstract}

\maketitle

\section{Introduction}

The determination of photospheric parameters of very hot central stars with
effective temperatures around \Teff=100\,000\,K is most comprehensively achieved
from quantitative NLTE model atmosphere analyses of optical \emph{and}
ultraviolet spectra. In particular, abundances of most metals can only be found
from UV observations because the strongest spectral lines from the highly
ionised species are located in this wavelength band. Temperature determinations
from optical spectra alone suffer from the absence of neutral helium lines,
preventing the use of the {He\,}{\sc i}/{He\,}{\sc ii} ionisation balance, which
is a common tool applied to cooler central stars. In addition, the Balmer line
problem at these high temperatures is still unsolved, i.e., models still cannot
achieve a fit to all Balmer lines of a particular object at a unique value for
\Teff. However, metal lines from different ionisation stages are often detectable 
in UV spectra, allowing precise constraints on \Teff.

In this contribution we report on our progress in this field since our review
presented on the last IAU planetary nebulae symposium \citep{W03}.  We first
present new results for hydrogen-rich central stars and then turn to the
hydrogen-deficient PG1159 stars. These two groups represent separate post-AGB
evolutionary channels, the first representing canonical post-AGB evolutionary
theory, and the second probably being the outcome of a late helium-shell
flash. PG1159 stars are thought to be the progeny of Wolf-Rayet type central
stars ([WC]), which are dealt with in the contribution by \cite{Ha05} in these
proceedings.

\section{Hydrogen-rich central stars}

As described in detail by \cite{W03}, we have taken high-resolution UV spectra
of eight very hot central stars with the STIS spectrograph aboard HST:
NGC\,1360, NGC\,1535, NGC\,4361, NGC\,6853, NGC\,7293, Abell~36, LSS\,1362, and
LS\,V$+$4621. From this sample, NGC\,1535 clearly stands out because of very
strong P~Cygni line profiles. It was subject to a separate study because, in
contrast to the other objects, expanding model atmospheres have to be used for
its analysis. The results were presented by \cite{Ko04}. In essence,
temperature, gravity, and wind parameters determined by previous analyses were
confirmed and a contemporary analysis performed by \cite{HB04} with a different
model code arrives at similar results.

\begin{figure}
\includegraphics[width=\columnwidth]{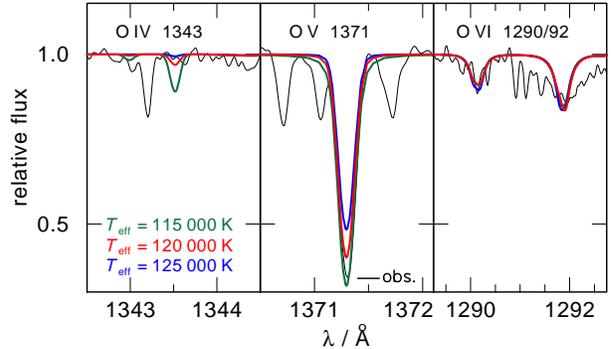} \caption{Details from a
high-resolution spectrum of NGC\,7293 (thin line) taken with HST/STIS showing
absorption lines from
three ionisation stages of oxygen, allowing to constrain \Teff. Overplotted are
three models with different \Teff\ as given in the left panel. {\bf Left:} The
{O\,}{\sc iv} line is not observed in the spectrum, excluding a temperature as
low as 115\,000\,K. Only the high \Teff\ models produce weak, undetectable
profiles. {\bf Middle:} The core of the observed {O\,}{\sc v} line is bracketed
by the models with 115\,000\,K and 120\,000\,K, excluding a temperature as high
as 125\,000\,K. {\bf Right:} The {O\,}{\sc vi} lines are matched satisfactorily,
however, they are insensitive against changes in \Teff\ in this parameter range.
} \label{fig_ngc7293_hst}
\end{figure}

From the other objects in our sample we performed a NLTE analysis with our {\sc
TMAP} code which computes line blanketed models in radiative and hydrostatic
equilibrium \cite{W03a}. We determined temperatures and gravities as well as
abundances of the CNO elements. When available, we have also utilized new far-UV
spectra taken with FUSE and optical spectra taken at different observatories
(Calar Alto 3.5m, Siding Spring 2.3m, Hobby-Eberly Telescope 9.2m). More
details were presented by \cite{Ho05} and \cite{Tr05}. As an example for the fit
procedure we display in Fig.\,\ref{fig_ngc7293_hst} details from the HST UV
spectrum of NGC\,7293, showing lines from oxygen in three ionisation stages
compared to models with different \Teff. It shows that temperatures can be fixed
with an accuracy of the order of 5\%. Our new, still preliminary results for \Teff\
and \logg\ as compared to previous, mostly optical, analyses, can be summarized
as follows. All objects, except NGC\,1360, are hotter than before, roughly by
10\,000--15\,000\,K. The most extreme case is NGC\,4361, whose \Teff\
``increased'' from 82\,000\,K to 126\,000\,K. We also correct for the gravities,
on the average by 0.3~dex, in either direction depending from star to star. By
comparison with theoretical evolutionary tracks we derived stellar masses. They
range between 0.55 and 0.65~M$_\odot$ and the mean mass is 0.60~M$_\odot$, in
agreement with the mean mass of white dwarfs. We find essentially solar abundances for the CNO elements in 5 of the
7 stars. This points at ineffective or even lacking
2nd and 3rd dredge-up events in the preceding RG and AGB phases, which can be
explained by the relatively low mass of the stars in our sample. According to
the initial-final-mass relation of \citep{Wei00}, our most massive post-AGB
remnant (0.65~M$_\odot$) had a main-sequence mass of 3~M$_\odot$.

Two objects definitely show non-solar abundance patterns. The first is
LS\,V$+$4621 (Sh2-216). It was previously known that this is a white dwarf
central star with a helium-deficiency (0.1 solar) explained by gravitational
settling \cite{Na99}. We find \Teff=93\,000\,K and \logg=6.9 and, in line with
the He-deficiency, a depletion of CNO elements between 1--2~dex. The second case
is NGC\,4361. It was realized that this is a halo-PN \citep{To90} so we expected
to find subsolar metal abundances. Indeed, it is obvious already on first sight,
that the iron lines are much weaker than in the other central stars and we find
that N is subsolar by 1~dex. However, O is found to be solar and, even more
surprising, the abundance of C is 20 times solar. This abundance pattern is
similar to that found for another Pop.\,II object, namely K\,648, the central
star of the planetary nebula Ps1 in the globular cluster M15 \citep{Ra02}. It
seems possible that carbon is enriched by $^{12}$C dredged up from the stellar
C/O core.

Our analysis of iron-group lines in the HST and FUSE spectra is still on-going
\citep{Ho05}. We often see iron lines of two ionisation stages in one
object. This will serve as an additional check for \Teff\ and we will determine
abundances of these heavy metals.

\begin{figure}
\includegraphics[width=\textwidth]{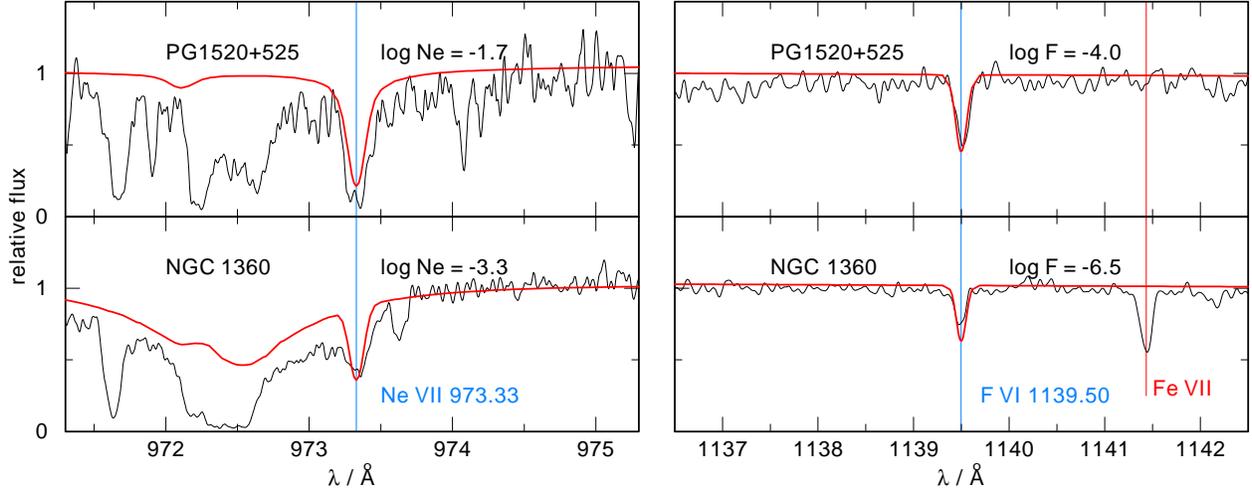} \caption{ Discovery of a
neon line (left panels) and a fluorine line (right panels) in the
hydrogen-deficient PG1159-type central star PG\,1520+525 (top panels) and in the
hydrogen-normal central star of NGC\,1360 (bottom panels). The neon and fluorine
abundances in the PG1159 star (given as mass fractions in the panels) are
strongly enhanced, namely 20 times and 250 times solar, respectively, whereas
they are solar in NGC\,1360. Note the strong {Fe\,}{\sc vii} line (not included
in the models) at 1141.4~\AA\ in NGC\,1360, which indicates a solar iron
abundance \citep{Ho05}. It is not detectable in the PG1159 star, probably due to
a subsolar iron abundance.} \label{fig_neon_fluorine}
\end{figure}

\begin{figure}
\includegraphics[width=\columnwidth]{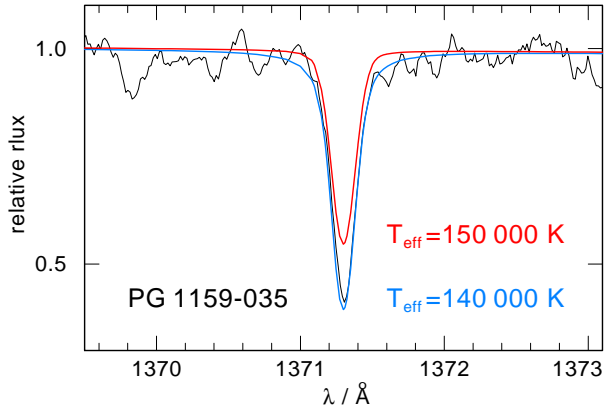} \caption{
Detail from a high-resolution spectrum of PG1159-035 taken with HST/STIS showing
the {O\,}{\sc v}~1371~\AA\ line. Overplotted are two model spectra with different
effective temperature. At \Teff=140\,000\,K the observed profile is almost
perfectly matched, whereas at \Teff=150\,000\,K the synthetic profile is too weak
in the line core.
} \label{fig_pg1159_hst}
\end{figure}

\begin{figure}
\includegraphics[width=\textwidth]{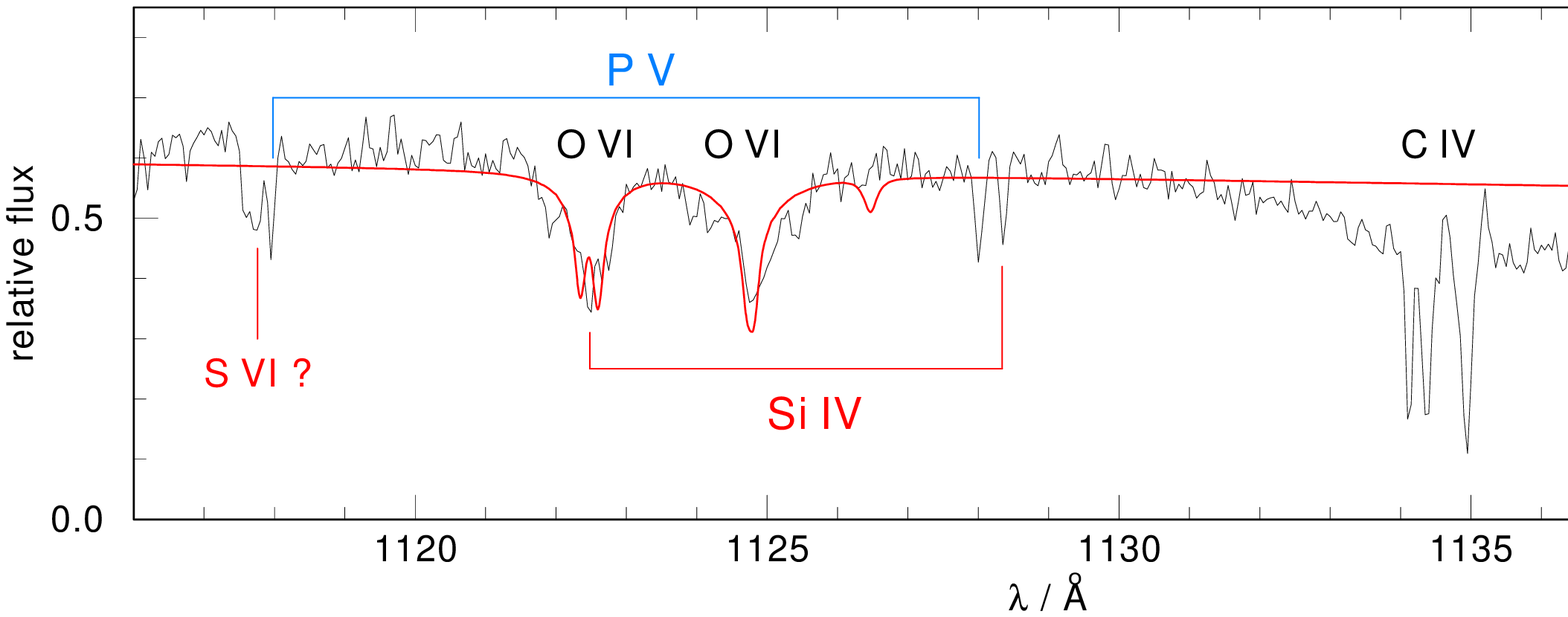} \caption{ First identification
of phosphorus and silicon in a PG1159 star (PG\,1424+535). Note also the
presence of a strong fluorine line and possibly a sulfur line (the {S\,}{\sc vi}
resonance doublet at 933/944~\AA\ is clearly detected). The P, Si, and S lines
are not yet included in the model, as well as the broad {C\,}{\sc iv} trough.  }
\label{fig_phosphorus}
\end{figure}

\section{Hydrogen-deficient central stars}

It now seems well established that the H-deficient central stars of spectral
type [WC] and PG1159 form an evolutionary sequence, although there still exists
an embarrassing systematic difference in the He/C ratios between early and late
type [WC] stars \cite{Ha05}. There is mounting evidence that the surface
composition of [WC] and PG1159 stars is identical to that of the intershell
matter in the progenitor AGB stars. The observed relative abundances of He, C,
and O are consistent with this picture. It is further confirmed by the
iron-deficiency found in a number of [WC] and PG1159 stars \citep{W03} which can
be explained by neutron captures of iron in the $^{13}$C pocket where
s-processing occurs. To us it seems most likely that the H-deficiency of these
stars is the result of a late He-shell flash which caused ingestion and burning
of the H-rich envelope \cite{Her99}. But in this case one would expect to
observe systematic differences in the nebula properties of H-rich and
H-deficient central stars which, however, are not evident \citep{Go00}. Another
problem with the late He-shell flash scenario is the simultaneous occurrence of
C- and O-rich circumstellar dust about [WC] central stars, which is difficult to
understand. An alternative scenario to explain the origin of [WC] stars with
this double-dust chemistry is a merging event of the AGB progenitor with a
planet or low-mass star, causing extra envelope mixing \citep{DM02}.

Whatever the cause of hydrogen-envelope mixing is (late helium-shell flash or merging
event), results of new spectral analyses further corroborate the picture that
such mixing results in the exposure of intershell matter on the surface. These
are the detection of a high abundances of neon in fluorine in a large number of
PG1159 stars.

For a few of the brightest PG1159 stars a neon abundance of 2\% (by mass) could
be derived from the {Ne\,}{\sc vii}~3644~\AA\ line \cite{WR94}. We have extended
these analyses to fainter objects using one of the VLT 8m telescopes
\citep{W04}. In addition, we identified a hitherto unknown strong neon line in
FUSE spectra, {Ne\,}{\sc vii}~973~\AA, which allows to determine the neon
abundance in PG1159 stars in a wider \Teff-\logg\ range than before. We also
identified a new {Ne\,}{\sc vii} multiplet in the optical wavelength range
(3850--3910~\AA). In all cases we find Ne=2\%, in full agreement with the neon
intershell abundance in AGB stars \cite{Her99}. In the upper left panel of
Fig.\,\ref{fig_neon_fluorine} we show as an example the {Ne\,}{\sc vii}~973~\AA\
line in the FUSE spectrum of the PG1159-type central star PG1520+525. The line
is so strong that it is still easily seen in hydrogen-rich central stars with a solar
neon abundance level, which is 20 times lower. This is presented in the lower left
panel of Fig.\,\ref{fig_neon_fluorine} which displays the FUSE spectrum of
NGC\,1360. In early [WC] stars and in the most luminous PG1159 stars this neon
line exhibits a strong P~Cygni profile. Formerly this line has been erroneously
interpreted as the {C\,}{\sc iii}~977~\AA\ resonance line but it could not be
matched by models because they showed too weak profiles due to high
ionisation \cite{HB05}.

FUSE spectroscopy also allowed for the first time to identify fluorine in
post-AGB stars \citep{W05}. In the right panels of Fig.\,\ref{fig_neon_fluorine}
we show the identification of the {F\,}{\sc vi}~1139.5~\AA\ line in the two
central stars just discussed. We find a solar F abundance in all five analysed
H-rich central stars but a wide abundance spread in the eight PG1159 stars: from
solar up to 250 times solar. The F enrichment can be explained by efficient F
production in the intershell of AGB stars, being in quantitative agreement with
nucleosynthesis calculations \citep{Lu04}. The F production efficiency is
strongly mass-dependent and this might be the reason for the large abundance
spread observed.

We have continued to investigate the iron-deficiency in PG1159 stars. We
planned to obtain high-resolution HST/STIS spectra of the prototype PG1159+035
and the central stars of NGC\,7094 and Abell~78 to possibly detect a nickel
enrichment caused by the transformation of iron to nickel during the s-process. The
observing programme was performed, except for the Abell\,78 observation because of
the fatal STIS failure in 2004. The spectra are still being analyzed. 
As an example of the quality of these spectra, we display in Fig.\,\ref{fig_pg1159_hst} 
the {O\,}{\sc v}~1371~\AA\ line profile for PG1159-035; this illustrates that we
can fix \Teff\ on a few-percent accuracy level.

We will also work on other trace elements in PG1159 stars to see if their
abundances are compatible with stellar evolution models. In this respect, the
FUSE spectra are, again, a rich source of information. For instance, in
Fig.\,\ref{fig_phosphorus} we display a detail from the spectrum of PG1424+535,
a relatively cool and high-gravity PG1159 star (\Teff=110\,000\,K, \logg=7). It
is located on the WD cooling sequence and its nebula has probably long being
dispersed. We identify lines from silicon, sulfur, and phosphorus; however, their
abundances still remain to be determined \cite{Rei05}.

As a final remark, we note that the high-resolution spectroscopy performed with
the hot central stars reveals problems with the available atomic data. Many of
the {O\,}{\sc vi} lines and the newly identified optical {Ne\,}{\sc vii}
multiplet appear at wavelength positions clearly deviating from atomic line
lists. In a positive sense, these spectra can be used to improve our
knowledge about the term structure of highly ionised metals. On the other hand,
many lines detected in the UV spectra remain unidentified even today.

\begin{theacknowledgments}
Analysis of HST and FUSE data in T\"ubingen is supported by DLR and DFG under
grants 50\,OR\,0201 and We\,1312/30-1, respectively.
\end{theacknowledgments}

\end{document}